\documentclass[prl,twocolumn,aps,showpacs,10pt,floatfix,longbibliography]{revtex4-1}

\usepackage[pdftex]{graphicx}  
\usepackage{epstopdf}
\usepackage{verbatim}
\usepackage{color}
\usepackage{subfigure}
\usepackage{tabularx}
\usepackage{amsfonts}
\usepackage{amsmath}
\usepackage{wasysym}
\usepackage{bm}
\usepackage{hyperref}

\begin{document}
\title{Topological thermal Hall effect for topological excitations in spin liquid:\\
Emergent Lorentz force on the spinons}
\author{Yong Hao Gao$^{1}$}
\author{Gang Chen$^{2,1}$}
\email{gangchen.physics@gmail.com}
\affiliation{$^{1}$State Key Laboratory of Surface Physics and Department of Physics, 
Fudan University, Shanghai 200433, China}
\affiliation{$^{2}$Department of Physics and Center of Theoretical and Computational Physics,
The University of Hong Kong, Pokfulam Road, Hong Kong, China}

\date{\today}
    
\begin{abstract}
We study the origin of Lorentz force on the spinons in a U(1) spin liquid. 
We are partly inspired by the previous observation of gauge field correlation in 
the pairwise spin correlation using the neutron scattering measurement by
P.A. Lee an N. Nagaosa [PhysRevB 87,064423(2013)] when 
the Dzyaloshinskii-Moriya interaction intertwines with the lattice geometry. 
We extend this observation to the Lorentz force that exerts 
on the (neutral) spinons. The external magnetic field, that polarizes the   
spins, effectively generates an internal U(1) gauge flux for the spinons    
and twists the spinon motion through the Dzyaloshinskii-Moriya interaction. 
Such a mechanism for the emergent Lorentz force differs fundamentally from  
the induction of the internal U(1) gauge flux in the weak Mott insulating 
regime from the charge fluctuations. We apply this understanding to the 
specific case of spinon metals on the kagome lattice. Our suggestion of  
emergent Lorentz force generation and the resulting topological thermal 
Hall effect may apply broadly to other non-centrosymmetric spin liquids 
with Dzyaloshinskii-Moriya interaction. We discuss the relevance with 
the thermal Hall transport in kagome materials volborthite and kapellasite.
\end{abstract}

\maketitle

Quantum spin liquid (QSL) is an exotic quantum state of matter in which 
spins are highly entangled quantum mechanically and remain disordered 
down to zero temperature~\cite{Balents2010,Lee2008,RevModPhys.89.025003}. 
Experimental identification of QSLs is of fundamental importance for our 
understanding of quantum matter. Thermal transport represents 
one sensitive experimental probe to unveil the nature of low-energy 
itinerant excitations, because other degrees of freedom, 
such as nuclear spins and defects, do not carry nor transport heat. 
Any heat current in a Mott insulator must be carried by the emergent and 
neutral quasiparticles~\cite{Nphys,Matsuda2010}. In the QSL regime, 
the deconfined spinons transport heat in the same way that the physical 
electrons carry charge in an electrical conductor. However, a major 
difficulty is that other excitations, 
most notably phonons, may get involved in the longitudinal thermal 
conductivity~\cite{PhysRevLett.120.117204,PhysRevB.92.094408,
PhysRevB.89.094403,PhysRevLett.121.147201,PhysRevX.8.031032,
Kasahara2018,PhysRevB.96.081111,PhysRevLett.120.067202,PhysRevLett.120.217205}.  
The quantitative contribution of spin excitations may be difficult 
to be extracted from the total longitudinal thermal conductivity 
due to the spin-phonon interaction, which is suggested 
to be present in many materials, especially in the ones with strong 
spin-orbit coupling. Thus, thermal Hall effect may be a more suitable 
probe to unveil the exotic excitations in QSLs since 
phonons do not usually contribute to thermal Hall transport.

\begin{figure}[b] 
\centering
\includegraphics[width=7.2cm]{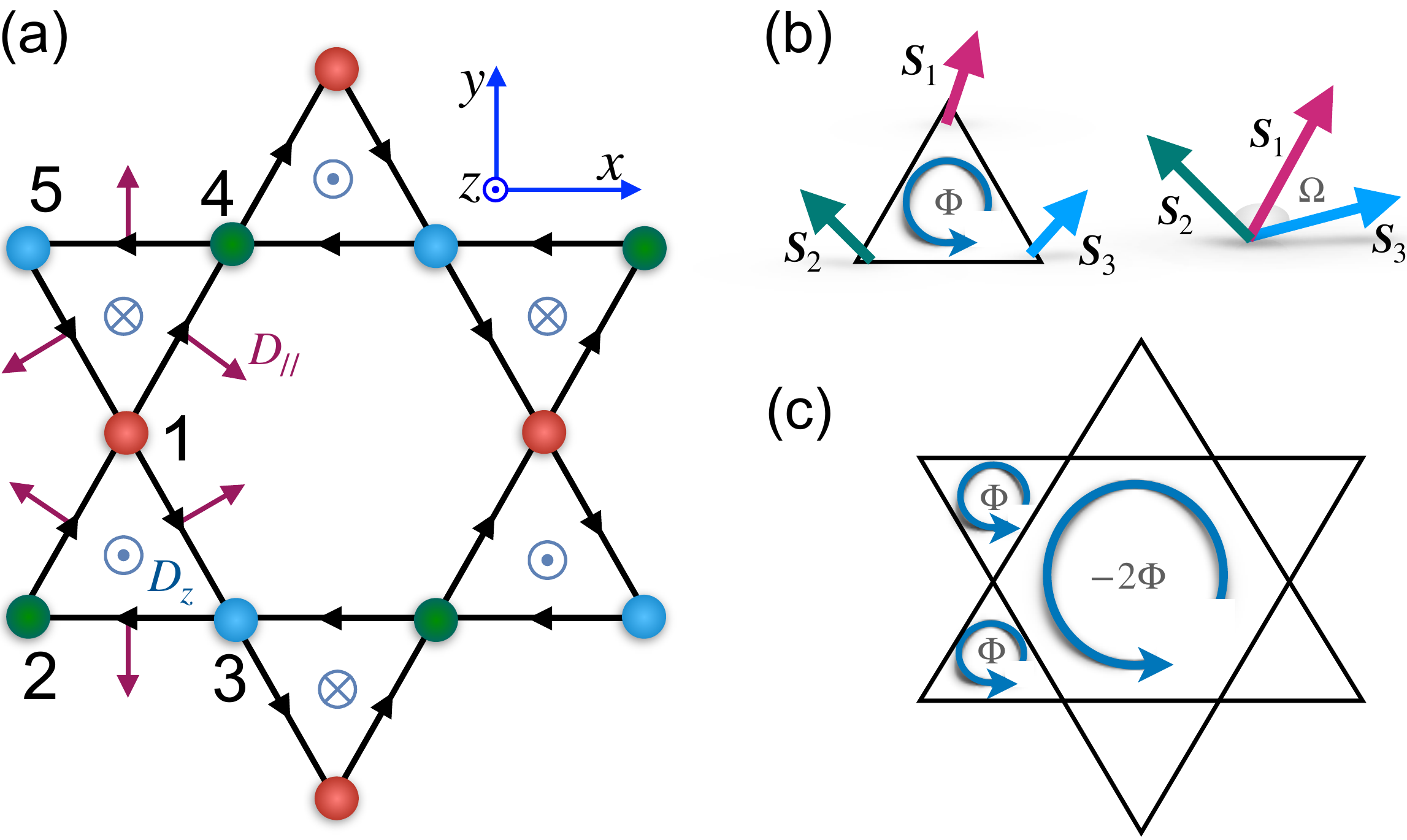}
\caption{(a) Symmetry allowed Dzyaloshinskii-Moriya interactions between 
first neighbors on the kagome lattice, where $D_z$ ($D_{\parallel}$) 
is the $z$ (in-plane) component. The black arrows on the bonds specify 
the order of the cross product $\boldsymbol{S_i}\times\boldsymbol{S_j}$. 
The sublattices are labelled by colors.  (b) Schematic view 
of scalar spin chirality for a non-collinear spin configuration, where $\Phi$  
is the corresponding gauge flux through the plaquette and $\Omega$ is 
the solid angle subtended by the three spins. (c) Internal U(1) flux distribution 
induced on the kagome lattice.}
\label{fig1}
\end{figure}

There are {\sl three ways} that thermal Hall effect may become signicant 
in a QSL. First, if the QSL is a two-dimensional chiral spin liquid, 
there would be chiral edge states that contribute a quantized thermal
Hall response. Second, if the external magnetic field comes to modify
the spinon bands such that the reconstructed spinon band 
develops edge states, the system would produce a quantized thermal 
Hall effect. A well-known example is the quantized thermal Hall effect 
in the Kitaev model~\cite{KITAEV20062} where the external field 
generates a Chern band for majorana spinons via high-order perturbations. 
This case may be not quite distinct from the first one except the first 
one is already a chiral spin liquid without magnetic field.
The third case is when the gauge field of the QSLs is continuous. This 
includes, for example, spinon Fermi surface U(1) QSL~\cite{PhysRevLett.95.036403,PhysRevB.72.045105,
PhysRevB.73.155115,PhysRevB.96.054445,PhysRevB.96.075105,PhysRevLett.121.046401}, 
U(1) Dirac QSL~\cite{PhysRevLett.98.117205,PhysRevB.77.224413,PhysRevB.72.104404}, and 
pyrochlore ice U(1) QSL~\cite{PhysRevB.69.064404,PhysRevB.86.104412,
PhysRevLett.108.037202,PhysRevLett.98.157204}. 
For the spinon Fermi surface U(1) QSL that was proposed    
for the weak Mott insulating organic materials $\kappa$-(ET)$_2$Cu$_2$(CN)$_3$
and EtMe$_3$Sb[Pd(dmit)$_2$]$_2$, it was 
suggested~\cite{PhysRevB.51.1922,PhysRevB.73.155115} that the external   
magnetic field could induce an internal U(1) gauge flux through the 
strong charge fluctuation or the four-spin ring exchange (due to the 
proximity to a Mott transition)~\cite{PhysRevLett.95.036403}. From this mechanism, 
the neutral spinons could experience the external field and contribute to the
thermal Hall effect~\cite{PhysRevLett.104.066403}, and a fundamentally 
different mechanism is required to understand the thermal Hall effects 
in this regime. Apparently, thermal Hall effects have been observed in 
the kagome magnets volborthite Cu$_3$V$_2$O$_7$(OH)$_2\cdot2$H$_2$O~\cite{Watanabe8653} 
and kapellasite CaCu$_3$(OH)$_6$Cl$_2\cdot$0.6H$_2$O~\cite{PhysRevLett.121.097203}, 
and the pyrochlore spin ice Tb$_2$Ti$_2$O$_7$~\cite{Hirschberger106}. 
In this Letter, we develop a theory of the topological thermal Hall effect
(TTHE) for U(1) QSLs with spinon Fermi surfaces in the strong Mott regime.  
We will explain the emergent Lorentz force generation and TTHE 
for the pyrochlore U(1) QSL in a forthcoming paper~\cite{1904}. 
In the end of this Letter, we discuss the open questions in this topic.

In the strong Mott insulating U(1) QSLs, the spinons carry emergent 
U(1) gauge charges and are minimally coupled to the U(1) gauge    
field as the spinons hop on the lattice. To twist the spinon 
motion, the external magnetic field has to influence the internal 
U(1) gauge field and then indirectly impacts on the spinon motion. 
In the strong Mott regime, the magnetic field couples to the spin
through the usual Zeeman coupling. The internal U(1) gauge flux 
is related to the scalar spin chirality, ${{\boldsymbol S}_i \cdot 
({\boldsymbol S}_j \times {\boldsymbol S}_k )}$, that involves three 
spins~\cite{PhysRevB.46.5621,PhysRevB.39.11413,RevModPhys.78.17}. 
It is not obvious how the linear Zeeman coupling enters  
to modify the three-spin scalar chirality in a disordered system, 
although both terms break the time reversal. A crucial observation 
was made by Patrick Lee and Naoto Nagaosa in the proposal~\cite{PhysRevB.87.064423} 
of detecting gauge fields or scalar spin chirality 
fluctuations using neutron scattering. They noticed that, with 
Dzyaloshinskii-Moriya interaction, the $S^z$-$S^z$
correlator contains a piece of the correlator of scalar spin chirality. 
Although their observation was originally made for neutron scattering,
it establishes the microscopic link between the Zeeman coupling
and the scalar spin chirality. In the following, we implement
this observation to understand the TTHE in QSLs.

In Mott insulators where the bond centers are not inversion centers, 
the Dzyaloshinskii-Moriya interaction is generally allowed~\cite{DZYALOSHINSKY1958241,PhysRev.120.91}. 
This is a relativistic effect and is more important in the strong  
spin-orbit-coupled systems such as the hyperkagome material Na$_4$Ir$_3$O$_8$~\cite{PhysRevB.78.094403}. 
A representative spin model in the strong Mott insulator has the form, 
\begin{equation}
H=\sum_{ i,j} J_{ij}\boldsymbol{S_i}\cdot\boldsymbol{S_j}
  +\sum_{ i,j }\boldsymbol{D}_{ij}\cdot\boldsymbol{S_i}\times\boldsymbol{S_j}
  - \sum_i B S^z_i,
  \label{spinmodel}
\end{equation}
where the direction of $\boldsymbol{D}_{ij}$ is determined by the lattice
symmetry from the Moriya's rule~\cite{PhysRev.120.91}, and the field is 
applied along $z$ direction. For the kagome lattice that is used below as 
an example to illustrate our thought, the Dzyaloshinskii-Moriya vector for 
nearest neighbors can have two components~\cite{PhysRevB.66.014422,PhysRevB.78.140405} 
with one normal to the kagome plane and the other in the kagome plane (see Fig.~\ref{fig1}(a)). 
This Hamiltonian with variant exchange couplings on neighboring bonds 
has been proposed for several kagome materials where spinon Fermi 
surface QSLs were suggested for some materials~\cite{PhysRevB.82.104434,Watanabe8653}. 
It has been estimated that the out-of-plane Dzyaloshinskii-Moriya term 
($D_z$) is about $8\%$ of the nearest-neighbor Heisenberg exchange for 
herbertsmithite~\cite{PhysRevLett.101.026405}. Our purpose is  
not to solve for the ground state of a specific Hamiltonian. 
We assume that the system stabilizes a U(1) QSL with a spinon 
Fermi surface and explain how the spinons acquire an emergent
Lorentz force from the Dzyaloshinskii-Moriya interaction.

For the spinon Fermi surface U(1) QSL, the spinon-gauge coupling is
described by the following Lagrangian, 
\begin{eqnarray}
\mathcal{L} &=& \sum_{i}f_{{i}\sigma}^{\dagger}
(\partial_{\tau}^{} - ia_{i}^{0}-\mu)f^{}_{{i}\sigma}
-\sum_{\langle ij \rangle } t\, e^{ia_{ij}}
f_{i\sigma}^{\dagger}f_{j\sigma}^{}
\nonumber \\
&& + \int_{dr}  \sum_{\mu} \frac{1}{g} (  \epsilon_{\mu\nu\lambda} \partial_{\nu} a_{\lambda} )^2,
\label{lag}
\end{eqnarray}
where the first line describes the spinon hopping on a kagome lattice 
and minimally coupled to the dynamical U(1) gauge field ${\boldsymbol a}$,
and the second line describes the fluctuation of ${\boldsymbol a}$.
The combined effect of the Dzyaloshinskii-Moriya interaction and 
Zeeman coupling has not been included at this stage. 
The connection between the emergent spinon-gauge variables and the 
spin variables is established from the usual Abrikovsov fermion construction 
with ${{\boldsymbol S}_i \equiv \frac{1}{2} f^\dagger_{i\alpha} 
\boldsymbol{\sigma}_{\alpha\beta}^{} f^{}_{i\beta}}$ (${\alpha,\beta = \uparrow,\downarrow}$) 
and the Hilbert space constraint ${ \sum_{\sigma} f^\dagger_{i\sigma} f^{}_{i\sigma} 
\equiv 1} $. As a standard procedure, 
the above spin-gauge coupling can be readily obtained by introducing 
the gauge fluctuation to the mean-field ansatz that generates the
spinon Fermi surface state~\cite{PhysRevLett.95.036403,PhysRevB.72.045105,PhysRevB.96.054445}. 
From Elitzur's theorem, only gauge invariant 
variables are related to the physical spins. The scalar spin chirality is 
related to the emergent U(1) gauge flux $\Phi$ via 
(see Fig.~\ref{fig1}(b))~\cite{PhysRevB.87.064423,PhysRevB.39.11413}
\begin{eqnarray}
\sin\Phi=\frac{1}{2}\boldsymbol{S}_1\cdot\boldsymbol{S}_2\times\boldsymbol{S}_3,
\end{eqnarray}
where the plaquette for the flux is defined by connecting the three spins.

For this U(1) QSL, we show below that the Dzyaloshinskii-Moriya interaction  
and Zeeman coupling together could generate a gauge flux distribution on the 
kagome lattice. The Dzyaloshinskii-Moriya interaction in the spin Hamiltonian
generates a finite vector spin chirality $\langle \boldsymbol{S}_i\times\boldsymbol{S}_j \rangle$.
This immediately suggests the linear relationship between the scalar spin chirality
and the vector spin operator. The Zeeman coupling generates a finite spin 
polarization. Thus, we have a finite scalar spin chirality on the lattice. 
To be specific, for the kagome lattice in Fig.~\ref{fig1}, we have 
\begin{equation}
\begin{gathered}
  \langle\boldsymbol{S}_2\times\boldsymbol{S}_3\rangle
= \langle\boldsymbol{S}_4\times\boldsymbol{S}_5\rangle
= \lambda \boldsymbol{D}_{23}=\lambda \boldsymbol{D}_{45} , 
\end{gathered}
\end{equation}
where $\lambda$ is a proportionality constant with ${\lambda\sim \mathcal{O} (J^{-1})}$,
and $J$ would be the largest exchange coupling.  
It is ready to see the linear relation between 
${\boldsymbol{S}_i\cdot\boldsymbol{S}_j\times\boldsymbol{S}_k}$ 
and ${\boldsymbol{S}_i\cdot\boldsymbol{D}_{jk}}$. 
Since we apply the magnetic field along $z$ direction, 
one then establishes $\langle \sin\Phi\rangle\simeq  
\frac{1}{2}\lambda D_z\langle S^z\rangle = \frac{1}{2}\lambda D_z
\chi B$, where $\Phi$ 
is the flux defined on the elementary triangular plaquette 
of the kagome lattice and $\chi$ is the magnetic susceptibility.
For the spinon Fermi surface QSL, $\chi$ is a constant.  
From the signs of the Dzyaloshinskii-Moriya interaction, we
conclude that the induced internal U(1) fluxes by the external magnetic 
field on both the up triangle and the down triangle are equal and denoted
as $\Phi$. The orientation of the flux loop is depicted in Fig.~\ref{fig1}(c).
Moreover, the flux through the hexagon is determined by fluxes in 
its six neighboring triangles. One can readily verify it equals 
$-2\Phi$ if adopting the anticlockwise loop convention in Fig.~\ref{fig1}(c).

We have demonstrated that the external magnetic field  
induces an internal U(1) gauge flux through the combination 
of Zeeman coupling and Dzyaloshinskii-Moriya interaction for 
a strong Mott insulator QSL. This U(1) gauge flux generation differs 
fundamentally from the induction of the internal U(1) gauge  
flux from the charge fluctuations in a weak Mott insulator QSL. 
The induced internal flux for strong Mott insulators from our 
mechanism depends on the direction of the Dzyaloshinskii-Moriya interaction
and is thus tied to the lattice geometry or symmetry. In contrast, for 
weak Mott insulators where the degrees of freedom are basically electrons,
the Lorentz coupling induced flux is always uniform and does not depend 
on the lattice geometry. Via our mechanism, 
the spinon motion in strong Mott insulators will be twisted by the 
induced internal U(1) gauge flux. This emergent Lorentz force on 
the spinons generates a topological thermal Hall effect (TTHE) of the spinons.
Our notion of ``TTHE'' is analogous to the ``topological Hall effect'' 
for itinerant magnets with non-collinear 
spin configurations such as skyrmion lattices that create a finite 
scalar spin chirality and effective U(1) gauge flux for the conduction electrons~\cite{PhysRevLett.83.3737,Gobel2018}. We recently learned the notion of TTHE was first
introduced in the thermal Hall transport for magnons~\cite{PhysRevB.95.014422}
in ordered magnets with non-trivial magnon band structure.

\begin{figure}[t] 
\centering
\includegraphics[width=8.2cm]{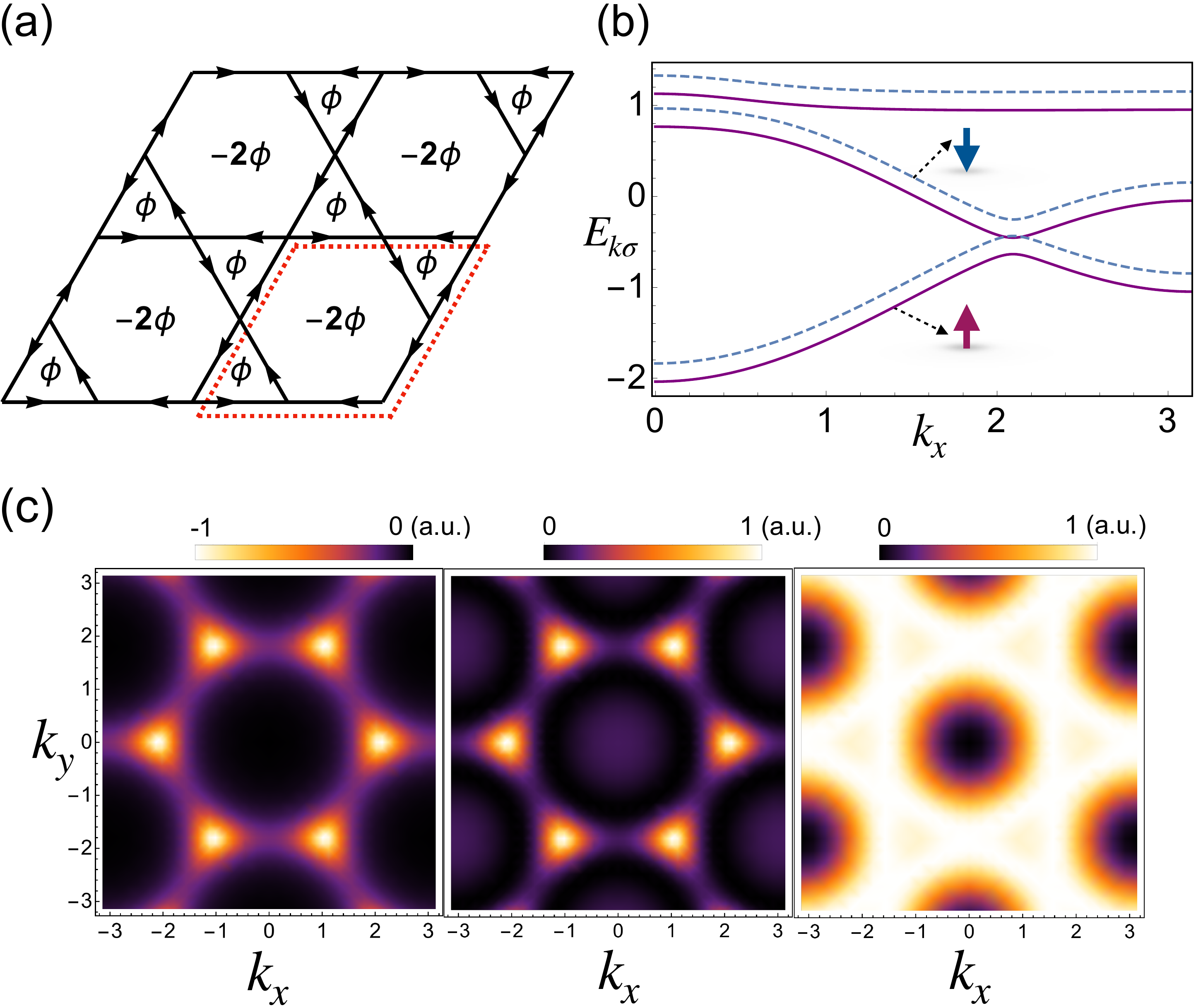} 
\caption{(a) The kagome lattice with U(1) gauge flux induced 
by external field through Dzyaloshinskii-Moriya interaction. 
The arrows on the bonds indicate the sign of the phase factor 
$e^{i\phi/3}$ and the flux through triangles and hexagons  
are $\phi$ and $-2\phi$, respectively. The area enclosed by 
the red dotted line is the unit cell of the kagome lattice. 
(b) Spinon bands for ${\phi=\pi/10}$ and the solid (dashed) 
lines are the bands for spin-$\uparrow$ (-$\downarrow$) 
spinons. (c) Density plot of the Berry curvature
 $\Omega_{n\boldsymbol{k}\sigma}$ of the lowest,  
 middle and highest bands for spin-$\uparrow$ 
 spinons, where we set 
 ${k_BT/t= 1}$ and ${\phi=\pi/3}$.}	
\label{fig2}
\end{figure}

In the standard linear response theory to an external magnetic field, 
the field enters as a perturbation. 
For the temperature gradient, however, the Hamiltonian stays invariant 
while the distribution function $e^{-\beta H }$ is modified \cite{PhysRevB.89.054420}, 
thus the theoretical treatment requires some care. This difficulty is overcome by the 
introduction of a fictitious pseudogravitational potential as shown by Luttinger \cite{PhysRev.135.A1505}. 
The temperature gradient is defined by ${T(\boldsymbol{r})=T_0[1-\eta(\boldsymbol{r})]}$ 
with a constant $T_0$ and a space-dependent small parameter $\eta(\boldsymbol{r})$, 
that can be regarded as a space-dependent prefactor to the Hamiltonian, 
${e^{-H/[k_BT(\boldsymbol{r})]}\simeq e^{-(1+\eta(\boldsymbol{r}))H/(k_BT_0)}}$.
Then, $\eta(\boldsymbol{r})H$ is regarded as a perturbation to the Hamiltonian 
from the temperature gradient. We can incorporate the temperature gradient into 
the Hamiltonian as a perturbation by using the psedogravitional potential. Further, we assume $\eta(\boldsymbol{r})$ to be linear in the position and expand the response in terms of $\nabla\eta(\boldsymbol{r})$ since we are interested in the linear response. The energy current density can then be derived as follows,
${j_{\mu}^{E}(\boldsymbol{r})=j_{0\mu}^{E}(\boldsymbol{r})+j_{1\mu}^{E}(\boldsymbol{r})}$,
where $j_{0\mu}^{E}(\boldsymbol{r})$ is independent of $\nabla\eta(\boldsymbol{r})$ 
and $j_{1\mu}^{E}(\boldsymbol{r})$ is linear in $\nabla\eta(\boldsymbol{r})$. 
They both contribute to the thermal transport coefficients. Ref.~\onlinecite{PhysRevB.89.054420} 
derived the thermal Hall conductivity for a noninteracting spinless boson Hamiltonian 
and was often used in the literature~\cite{PhysRevLett.106.197202,PhysRevB.84.184406}. 
Since we are dealing with 
fermionic spinons, so we adopt the result from Ref.~\onlinecite{PhysRevLett.107.236601} 
where a thermal Hall conductivity formula for a general noninteracting fermionic system 
with a nonzero chemical potential $\mu$ was obtained as
\begin{equation}
\label{thermalcon}
	\kappa_{xy}=-\frac{1}{T}\int d\epsilon(\epsilon-\mu)^2\frac{\partial f(\epsilon,\mu,T)}{\partial \epsilon}\sigma_{xy}(\epsilon).
\end{equation}	
Here ${f(\epsilon,\mu,T)=1/[e^{\beta(\epsilon-\mu)}+1]}$ is the Fermi-Dirac distribution 
and ${\sigma_{xy}(\epsilon)=-1/\hbar\sum_{\boldsymbol{k},\sigma,\xi_{n,\boldsymbol{k}}<\epsilon}\Omega_{n,\boldsymbol{k},\sigma}}$ is the zero temperature anomalous Hall coefficient 
for a system with the chemical potential $\epsilon$.
$\Omega_{n\boldsymbol{k}\sigma}$ 
is the Berry curvature for the fermions and is defined as
${\Omega_{n\boldsymbol{k}\sigma}=-2 {\rm Im}\langle  
{\partial_{k_x} u_{n\boldsymbol{k}\sigma}}|{\partial_{\partial k_y} u_{n\boldsymbol{k}\sigma}}\rangle}$
with eigenstate $\vert u_{n\boldsymbol{k}\sigma} \rangle$ for band indexed by 
$n$ and the spin $\sigma$. Eq.~\eqref{thermalcon} indicates
that the thermal Hall conductivity is directly related to the Berry curvature 
in momentum space and a finite Berry curvature is necessarily required to 
generate $\kappa_{xy}$. We show below that the magnetic field induced internal 
U(1) gauge flux generates a finite Berry curvature and use Eq.~\eqref{thermalcon} 
as our basis to calculate thermal Hall conductivity for the spinon metal 
in a U(1) QSL.

To describe the TTHE in the spinon metal, we consider a 
mean-field Hamiltonian for the spinon metal in the external magnetic field 
without including the U(1) gauge fluctuations of Eq.~\eqref{lag},
$H_{\text{MF}} =
- \sum_{\langle ij \rangle} [t_{ij}f_{i\sigma}^{\dagger}f_{j\sigma}^{}  + h.c.]
- \mu\sum_{i }f_{i\sigma}^{\dagger}f^{}_{i\sigma}  
 - B\sum_{i,\alpha\beta} \frac{1}{2} f_{i\alpha}^{\dagger}{\sigma_{\alpha\beta}^{z}}f_{i\beta}$, 
where the chemical potential $\mu$ is introduced to impose the Hilbert space constraint
and the effect of the Dzyaloshinskii-Moriya interaction is not included here. This 
free-spinon mean-field Hamiltonian simply describes a QSL with a large spinon 
Fermi surface in the weak magnetic field. As we have explained above, the combination 
of the microscopic Dzyaloshinskii-Moriya interaction and Zeeman coupling induces an 
internal U(1) gauge flux distribution on the kagome plane. To capture this flux pattern
in Fig.~\ref{fig1}, we modify the spinon mean-field Hamiltonian by adding the U(1) gauge 
potential with
\begin{eqnarray}
H_{\text{MF}}[\phi] &=& -t\sum_{\langle ij \rangle} 
[e^{-i\phi/3}f_{i\sigma}^{\dagger}f_{j\sigma}^{} + h.c.]
- \mu\sum_{i }f_{i\sigma}^{\dagger}f^{}_{i\sigma}  \nonumber \\
&& 
- B\sum_{i,\alpha\beta}f_{i\alpha}^{\dagger}\frac{\sigma_{\alpha\beta}^{z}}{2}f_{i\beta},
\label{mft2}
\end{eqnarray}
where we have fixed the gauge by setting the U(1) gauge field ${\langle a_{ij}\rangle =\phi/3}$
for all the nearest-neighbor spinon hopping in the anticlockwise manner. 
The net flux in each unit cell is zero (see Fig.~\ref{fig2}(a)), so 
the translation symmetry of the spinons is not realized projectively.

Without the internal U(1) gauge flux, the spinon Hamiltonian $H_{\text{MF}}$
is real, and one can always choose the eigenvector $\vert u_{n\boldsymbol{k}\sigma} \rangle$ 
to be real unless there is a band degeneracy, which immediately 
gives ${\Omega_{n\boldsymbol{k}\sigma}=0}$. With the internal U(1) gauge flux, 
the spinon Hamiltonian in Eq.~\eqref{mft2} is complex and we expect a 
finite Berry curvature. Indeed as we plot in Fig.~\ref{fig2} for the specific
choices of fluxes, the internal U(1) gauge flux reconstructs the spinon 
bands and creates the Berry curvatures of the spinon bands. 
The induced flux eliminates the band touching at $\Gamma$ point between the 
upper two bands and the Dirac band touching ${\rm K}$ point between the lower 
two bands. The Zeeman coupling further splits the spinon bands with up and 
down spins. Berry curvatures are enhanced 
at ${\rm K}$ point for the lower two bands and along 
the Brillouin zone boundary for the highest bands.

\begin{figure}[t] 
\centering
\includegraphics[width=8.6cm]{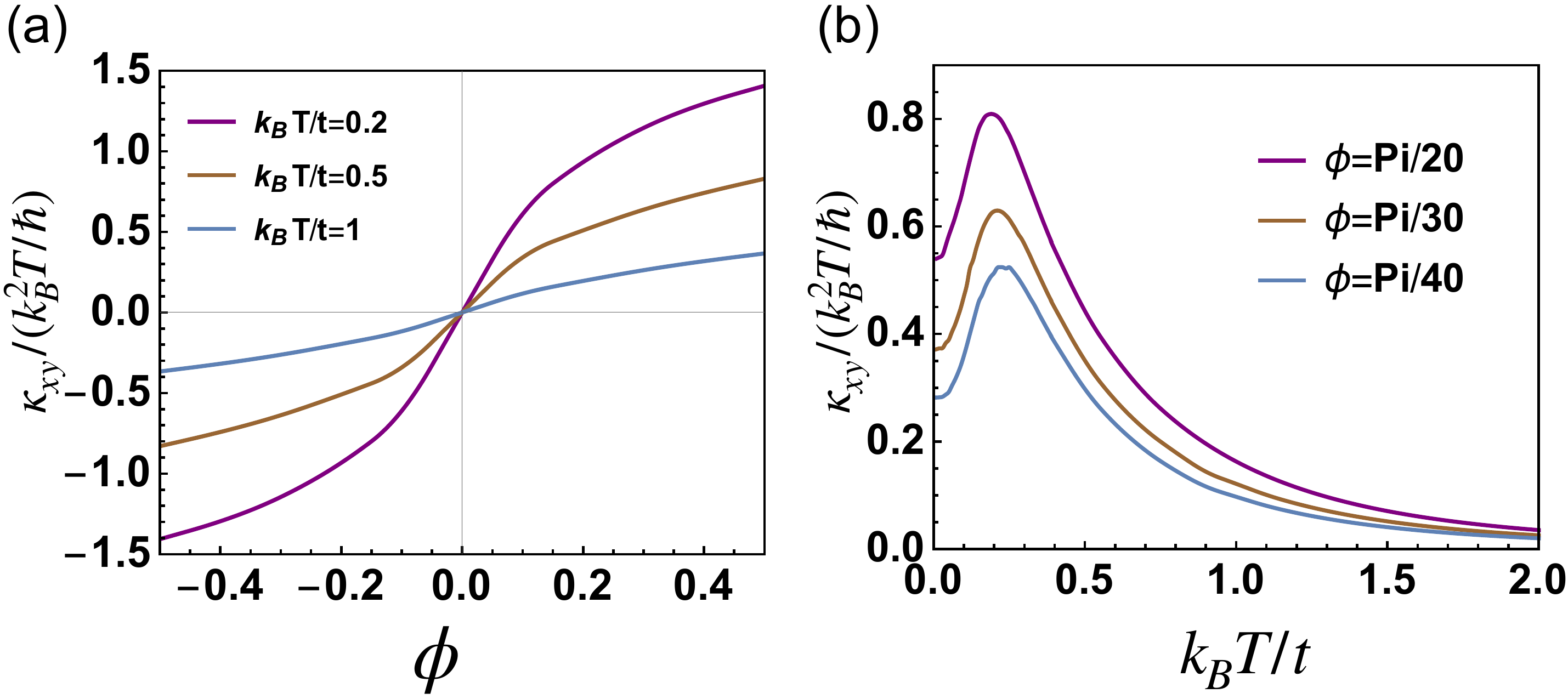}\\
\caption{(a) The dependence on the induced internal flux $\phi$ of the thermal Hall conductivity 
at several temperatures. (b) The thermal Hall conductivity as a function of temperature.}
\label{fig3}
\end{figure}

We calculate the thermal Hall conductivity for our TTHE based on 
the spinon mean-field Hamiltonian Eq.~\eqref{mft2} using the formula 
Eq.~\eqref{thermalcon} by varying the flux and the temperature. 
The results are depicted in Fig.~\ref{fig3}. The thermal Hall 
conductivity $\kappa_{xy}$ vanishes at zero flux ({\sl i.e.} at 
zero field) and increases monotonously 
with a finite flux $\phi$ in the zero flux limit. Due to the 
spinon Fermi surface, $\kappa_{xy}/T$ becomes a constant in the zero
temperature limit~\footnote{Yi Zhou, Private communications, Jan 2019}. 
The non-monotonic temperature dependence appears 
at finite temperatures. At very high temperatures, $\kappa_{xy}/T$ should 
certainly vanish because the spinons are almost equally populated and 
the summation of Berry curvatures of all bands vanishes, and moreover,
the magnetic susceptibility would become very small at high temperatures
and suppress the induced internal gauge flux. 
At very low temperatures, the spinon chemical potential decreases as $T$ increases.
In this limit, $\kappa_{xy}/T$ can be approximated as the summation of 
Berry curvature of spinon bands with energies below the chemical potential~\cite{Supple}.
As the chemical potential sits on the middle band, and the 
Berry curvatures of the lowest and middle bands are of opposite sign, 
the Berry curvature cancellation from two lowest bands becomes less, thus
we would expect an increase of $\kappa_{xy}/T$ as $T$ increases. 
This explains the non-monotonic temperature dependence.

\emph{Discussion}---In summary, we have proposed a physical mechanism of
the emergent Lorentz force on spinons and established the resulting TTHE 
in QSLs. We applied this understanding to the specific cases of spinon metals in kagome lattice and calculated the
TTHE. It offers a new perspective to understand the origin of thermal Hall effect of QSLs in strong Mott regime and can be related to the clear thermal Hall signal observed recently in kagome materials volborthite and kapellasite~\cite{Watanabe8653,PhysRevLett.121.097203}, 
since the main feature of the experimental $\kappa_{xy}$ in the QSL 
region (such as non-monotonic temperature dependence) are consistent with our theoretical result. 
The opposite signs of the thermal Hall conductivities in volborthite and kapellasite
could arise from the opposite signs of the Dzyaloshinskii-Moriya interaction that 
induces the internal U(1) fluxes with opposite signs. 
Our theory can apply broadly to other non-centrosymmetric QSLs with 
Dzyaloshinskii-Moriya interaction and QSLs with bosonic spinons.
Our understanding based on the emergent Lorentz force and/or the induced internal 
U(1) gauge flux through Dzyaloshinskii-Moriya interaction  
differs from the calculation using the bosonic spinon 
and Schwinger boson mean-field theory for gapped QSLs by 
Ref.~\cite{PhysRevLett.121.097203} for kagome kapellasite 
and more recently in Ref.~\cite{subir1812.08792} for the square lattice. 
In the Supplementary Material~\cite{Supple}, 
we further contrast our mechanism with the one from the strong charge 
fluctuation in the weak Mott regime.

Broadly speaking, thermal transport in Mott insulators is an interesting 
direction in quantum magnetism~\cite{PhysRevLett.104.066403}. 
In the high temperature paramagnet, the high
temperature series expansion can be applied. 
In the intermediate temperature regime where the correlation deleveps but 
there is no quasiparticle description yet, the thermal transport of these
``no-particles'' is an open subject in the field. The thermal transport
on the pyrochlore ice material Tb$_2$Ti$_2$O$_7$ remains to be understood. 
In the very low temperature, various quasiparticle descriptions may 
emerge. For ordered magnets, magnons would be the energy carriers. 
The study of magnon Berry curvature has proved successful in the 
thermal Hall study of pyrochlore ferromagnet Lu$_2$V$_2$O$_7$~\cite{Onose297}
and the kagome ferromagnet Cu(1,3-benzenedicarboxylate)~\cite{PhysRevLett.115.106603}. 
For QSLs, the quasiparticle description is given by the parton-gauge language.
Our current work about TTHE in QSL, that is based on the coupling between the 
spinon and the U(1) gauge field and is independent of the statistics 
of the spinons, elucidates the keen link between 
the emergent objects (such as the internal gauge 
flux, emergent Lorentz force and spinon Berry curvature) 
and the microscopic objects (such as the external magnetic field 
and Dzyaloshinskii-Moriya interactions)
and provides the microscopic understanding of TTHE.

\emph{Acknowledgments.}---This work is supported by the Ministry of Science and 
Technology of China with the Grant No. 2016YFA0301001, 2016YFA0300500, 2018YFGH000095
and by Grant funding from Hong Kong's Research Grants Council (GRF no.17303819).

\newpage

{Supplementary Material for ``Topological thermal Hall effect for topological excitations in spin liquid''}

\date{\today}

\section{Scalar spin chirality and the instantaneous gauge flux} \label{sec22}

In this Supplementary material, following the main text, we adopt the canonical  
Abrikovsov  fermion representation. We now consider three sites  around a plaquette labeled by $i,j,k$ in an anticlockwise manner and define $P_{ijk}=\langle\chi_{ij}\chi_{jk}\chi_{ki}\rangle$. $P_{ijk}$ can also be represented by $P_{ijk}=\langle f_{i\alpha}^{\dagger}f_{j\alpha}f_{j\beta}^{\dagger}f_{k\beta}f_{k\gamma}^{\dagger}f_{i\gamma}\rangle$ from the definition of $\chi_{ij}$. Further defining the operator $\hat{P}_{ijk}= f_{i\alpha}^{\dagger}f_{j\alpha}f_{j\beta}^{\dagger}f_{k\beta}f_{k\gamma}^{\dagger}f_{i\gamma}$, simple calculations \cite{PhysRevB.46.5621, RevModPhys.78.17}show that
\begin{equation} \label{eq5}
\hat{P}_{ijk}-\hat{P}_{ikj}=4 i \boldsymbol{S}_i\cdot\boldsymbol{S}_j\times\boldsymbol{S}_k.
\end{equation}
The quantity $\boldsymbol{S}_1\cdot\boldsymbol{S}_2\times\boldsymbol{S}_3$ is the so-called scalar spin chirality and is one of the key concepts in strong correlated physics. Ignoring amplitude fluctuations, one will have 
\begin{equation} \label{eq6}
\langle\hat{P}_{ijk}-\hat{P}_{ikj}\rangle=\chi_0^3\left(e^{i(\theta_{ij}+\theta_{jk}+\theta_{ki})}+c.c.\right).
\end{equation}
The combination of $\theta_{ij}$ in the exponent is the sum of the gauge field variables around a plaquette, which is the gauge invariant flux  through the plaquette. Combing Eq.~\eqref{eq5} and Eq.~\eqref{eq6}, we can see that in the QSL state, if $\Phi$ is the instantaneous gauge flux through a triangle formed by sites $1, 2$ and $3$ in a counterclockwise way, we have $\sin\Phi=\frac{1}{2}\boldsymbol{S}_1\cdot\boldsymbol{S}_2\times\boldsymbol{S}_3$, i.e., $\sin\Phi$ is one-half of the solid angle subtended by three spins $\boldsymbol{S}_i (i=1,2,3)$, as depicted in Fig. 1 (b) in the main text. In this sense, the fluctuations in the gauge field can be interpreted as fluctuations in the chirality through each plaquette.
For the kagome lattice with symmetry allowed Dzyaloshinskii-Moriya interactions (see Fig. 1 (a)), there is an important connection \cite{PhysRevB.87.064423} between $\boldsymbol{S}_i\times\boldsymbol{S}_j$ and $\boldsymbol{D}_{ij}$  due to the Dzyaloshinskii-Moriya term in the spin Hamiltonian, which gives
\begin{equation}
\begin{gathered}
\langle\boldsymbol{S}_2\times\boldsymbol{S}_3\rangle=\lambda \boldsymbol{D}_{23},\\
\langle\boldsymbol{S}_4\times\boldsymbol{S}_5\rangle=\lambda \boldsymbol{D}_{45}=\lambda\boldsymbol{D}_{23},
\end{gathered}
\end{equation}
where $\lambda$ is a constant estimated at $\lambda\sim1/J$. It is readily verify there exists a linear coupling between the spin chirality $\boldsymbol{S}_i\cdot\boldsymbol{S}_j\times\boldsymbol{S}_k$ and $\boldsymbol{S}_i\cdot\boldsymbol{D}_{jk}$. Averaging the total chirality through the two attached up and down triangles, the in-plane component $D_{\parallel}$ of the Dzyaloshinskii-Moriya vectors will be canceled out and one  can obtain $\langle \sin\Phi\rangle=1/2\lambda D_z\langle S^z\rangle$. A more formal proof about the above relations can be proceeded within the first order perturbation theory \cite{PhysRevB.87.064423}.

\section{Wiedemann-Franz Law in the zero temperature limit} 

Let us now verify that the thermal Hall conductivity formula \cite{PhysRevLett.107.236601} Eq. (5) in the main text for the fermion systems recovers the usual Wiedemann-Franz Law for a noninteracting system in the zero temperature limit. The derivative of the Fermi-Dirac distribution function $ \partial f(\epsilon,\mu,T)/\partial \epsilon$ indicates that the integral in Eq. (5) dominates around the Fermi energy. In the zero temperature limit, it represents a sharp peak and can be expanded as
\begin{equation}
\frac{\partial f(\epsilon,\mu,T)}{\partial \epsilon}=-\delta(\epsilon-\mu)-\frac{(\pi k_B T)^2}{6}\frac{d^2}{d \epsilon^2}\delta(\epsilon-\mu)+...
\end{equation}
Thus the thermal Hall conductivity is recast into
\begin{equation}
\kappa_{xy}=\frac{\pi^2 k_B^2T}{6}\int d\epsilon(\epsilon-\mu)^2\frac{d^2}{d \epsilon^2}\delta(\epsilon-\mu)\sigma_{xy}(\epsilon).
\end{equation} 
Using the relation $\delta''(x)=2\delta(x)/x^2$, one can easily obtain
\begin{equation}
\kappa_{xy}=\frac{\pi^2k_B^2T}{3}\sigma_{xy}(\mu).
\end{equation}
Remarkably, $\kappa_{xy}/T=\frac{\pi^2k_B^2}{3}\sigma_{xy}(\mu)$ suggests, in the zero temperature limit, that $\kappa_{xy}/T\neq0$ if $\sigma_{xy}(\mu)$ is non-vanishing.
In terms of the Berry curvature $\Omega_{n\boldsymbol{k}\sigma}$, one can further re-express the thermal Hall conductivity as
\begin{equation}
\kappa_{xy}=-\frac{\pi^2k_B^2T}{3\hbar}\sum_{\boldsymbol{k},\sigma,\xi_{n,\boldsymbol{k}}<\mu}\Omega_{n,\boldsymbol{k},\sigma},
\end{equation}
which directly indicates that a non-zero Berry curvature is necessarily needed to contribute to the thermal Hall conductivity.

\section{Mean-field free spinon Hamiltonian} 
\label{sec4}

To diagonalize the Hamiltonian, by performing the Fourier transform of the mean-field Hamiltonian Eq. (6) in the main text into momentum space, one can obtain  $H=\sum_{\boldsymbol{k},\alpha}\psi^{\dagger}_{\boldsymbol{k},\alpha}h_{\boldsymbol{k},\alpha}\psi_{\boldsymbol{k},\alpha}$, with basis $\psi_{\boldsymbol{k},\alpha}=(f^{u}_{k,\alpha},f^{v}_{k,\alpha},f^{w}_{k,\alpha})^T$ and the matrix
\begin{equation} \label{Hami}
\begin{split}
h_{\boldsymbol{k},\alpha}&=(-\mu-\alpha \frac{B}{2})I_3\\
&-2t\begin{pmatrix}
0& \cos k_1e^{-i\phi/3}& \cos k_3e^{i\phi/3}&\\
\cos k_1e^{i\phi/3}& 0& \cos k_2e^{-i\phi/3}&\\
\cos k_3e^{-i\phi/3}& \cos k_2e^{i\phi/3}& 0&\\
\end{pmatrix}
\end{split}
\end{equation}
where $u, v$ and $w$ are three sublattice indexes and $k_i=\boldsymbol{k}\cdot\boldsymbol{a_i}$, with $\boldsymbol{a_1}=(-1/2,-\sqrt{3}/2)$, $\boldsymbol{a_2}=(1,0)$ and $\boldsymbol{a_3}=(-1/2,\sqrt{3}/2)$
the nearest-neighbor vectors. $I_3$ is a $3\times3$ identy matrix and the lattice constant has set to be unit. The chemical potential $\mu$ is self-consistently calculated by the following equation
\begin{equation}
\frac{1}{3N}\sum_{n,\boldsymbol{k},\alpha}n_F(\frac{E_{n\boldsymbol{k}\alpha}}{k_BT})=1,
\end{equation}
which guarantees the proper Hilbert space and $E_{n\boldsymbol{k}\alpha}$ is the dispersion of the $n$-th band of spin-$\alpha$ spinons.

\section{Comparison with the weak Mott regime} 

For the U(1) QSL with the spinon Fermi surface in the weak Mott insulator, it was suggested 
that the external magnetic field could induce an uniform internal flux distribution through 
the strong charge fluctuation~\cite{PhysRevB.73.155115}. 
More precisely, it is through the 4-site ring exchange interaction. 
As the system that was studied is a triangular lattice, 
the induced flux is an uniform flux on each triangular plaquette.  
This induced U(1) gauge flux would twist the motion of the spinons and generate 
a thermal Hall effect for the spinons~\cite{PhysRevLett.104.066403}. Because of the uniform flux distribution, 
the translation symmetry of the spinons is realized projectively. Moreover, in 
principle, one would expect a quantum oscillation in the thermal Hall conductivity 
from the uniform flux induced by the external field~\cite{PhysRevB.73.155115}. 

In contrast, for our mechanism, the combination of the Dzyaloshinskii-Moriya interaction
adn the Zeeman coupling generates a staggered flux distribution. The net flux in a unit cell 
of the kagome lattice is zero such that the translation symmetry of the spinon is realized 
unprojectively. Thus, one would not expect the quantum oscillation phenomena for our mechanism. 

Finally, as we have mentioned in the main text, our mechanism can be well extended to other U(1) QSLs 
with Dzyaloshinskii-Moriya interactions. This includes both strong Mott regime and weak
Mott regime. For the ones in the weak Mott regime, our mechanism and the previous mechanism from
strong charge fluctuation would come together. In the strong Mott regime, however, the 
charge fluctuation is suppressed, and the previous mechanism can be neglected.

The QSLs in weak Mott regime that were proposed have a triangular lattice 
structure~\cite{PhysRevB.73.155115,PhysRevLett.95.036403}. 
If one can approximate the candidate organic materials as perfect or nearly perfect triangular lattice, 
the Dzyaloshinskii-Moriya interaction is absent, so one does not need to invoke our mechanism 
for these systems.

\bibliography{Ref.bib}

\end{document}